\documentclass[11pt,a4paper]{article}

\usepackage{mathpple}
\usepackage{graphicx}
\usepackage{url}
\usepackage{paralist}

\usepackage{geometry}
\usepackage{fancyhdr}
\begin{document}

\thispagestyle{plain}

\noindent \textbf{Preprint of:}\\
D. K. Gramotnev, M. L. Mather and T. A. Nieminen\\
``Anomalous absorption of bulk shear sagittal acoustic waves
        in a layered structure with viscous fluid''\\
\textit{Ultrasonics} \textbf{41}, 197--205 (2003)

\hrulefill

\begin{center}

\textbf{\LARGE
Anomalous absorption of bulk shear sagittal acoustic waves
        in a layered structure with viscous fluid}

{\Large
Dmitri K. Gramotnev, Melissa L. Mather and Timo A. Nieminen}

Applied Optics Program, School of Physical and Chemical Sciences,
Queensland University of Technology, GPO Box 2434, Brisbane,
QLD 4001, Australia; e-mail: d.gramotnev@qut.edu.au

\vspace{1cm}

\begin{minipage}{0.8\columnwidth}
\section*{Abstract}
It is demonstrated theoretically that the absorptivity of bulk shear
sagittal waves by an ultra-thin layer of viscous fluid between
two different elastic media has a strong maximum (in some cases as
good as 100\%) at an optimal layer thickness. This thickness is
usually much smaller than the penetration depths and lengths of
transverse and longitudinal waves in the fluid. The angular dependencies
of the absorptivity are demonstrated to have significant and unusual
structure near critical angles of incidence. The effect
of non-Newtonian properties and non-uniformities of the fluid
layer on the absorptivity is also investigated. In particular, it is
shown that the absorption in a thin layer of viscous fluid is much
more sensitive to non-zero relaxation time(s) in the fluid layer than
the absorption at an isolated solid-fluid interface.
\end{minipage}

\end{center}

\section{Introduction}

One of the most important areas of application of
acoustic waves is related to non-destructive evaluation
and development of highly sensitive ultrasonic measurement
techniques and sensors. This is the main reason
for the intensive theoretical and experimental
research in the area of propagation and interaction of
various types of ultrasound waves near different kinds of
interfaces, layered structures, defects, structural imperfections,
etc. This includes theoretical and experimental
investigations of interaction of acoustic waves with
imperfect interfaces and sets of cracks or inclusions [1-5],
slipping contacts [3,4], sliding contacts [6,7], bonding
layers between two elastic media [8,9], propagation and
interaction of acoustic waves near solid-fluid interfaces
[10-17], etc. In these cases, various approaches based on
spring boundary conditions [1,2,4,5], transfer matrix
approach [8,9,19], equivalent boundary conditions
[8,9,20], Stroh formalism [6,7], etc. have been developed
and used.

   One of the most important areas of application of
acoustic waves is related to the development of new
highly sensitive viscosity sensors and techniques for
diagnostics of fluids and solid-fluid interfaces and fluid
layers [4,10-18]. Major advantages of using acoustic
waves for these applications are related to a possibility
of diagnostics of small amounts of fluids, investigation
of fluids in the immediate proximity of solid interfaces
(i.e., analysis of processes of interaction between a fluid
and a solid), possibility of sensor miniaturization and
increased sensitivity.

   In terms of development of new viscosity sensors and
fluid diagnostics techniques, sensitivity of measurements
is of prime importance. Surface acoustic waves usually
provide higher sensitivity than bulk waves, because
surface waves can easily be kept in contact with a fluid
for a long time as they propagate along a solid-fluid
interface. Therefore, ultrasonic viscosity sensors and
diagnostic techniques are usually based on surface
acoustic waves [12-14]. However, in some applications it
may be more convenient to use bulk acoustic waves.
Their main advantages over surface acoustic waves are
related to the absence of additional structures (such as
generating and receiving interdigital transducers or
periodic structures) at the solid-fluid interface, and much
higher spatial resolution of measurements. Therefore, if
we are able to improve sensitivity of viscosity sensors
using bulk acoustic waves, these devices will be much
more competitive as compared to surface acoustic wave
sensors. This can be achieved using the anomalous
absorption (AA) of acoustic waves by a thin layer of
viscous fluid enclosed between two solid media [21-28].
Moreover, application of AA for viscosity sensor design
may have additional advantages related to a possibility
of diagnostics of extremely small (less than $\approx 10^{-5}$ cm$^3$)
amounts of fluids, precise high frequency and low
frequency analysis of solid-fluid interfaces and areas of
fluid in the immediate proximity to the interface,
and acoustic diagnostics of frictional contacts and
lubricants. In addition, AA may be useful for the
development of new techniques for non-destructive evaluation.

   AA is characterized by a very strong increase in the
absorptivity of the incident wave by the fluid layer with
decreasing thickness of the layer (this is the reason for a
substantial increase in the sensitivity of viscosity sensors
using AA). The absorptivity reaches a maximum at an
optimal layer thickness, and then quickly goes to zero as
the layer thickness tends to zero [21-28]. The optimal
layer thickness is usually much smaller than penetration
depths and lengths of transverse and longitudinal waves
in the fluid [21-28]. Therefore, AA cannot be explained
by the attenuated total reflection or resonant interference
of bulk waves in the layered structure. It was explained
by a significant increase of the fluid velocity gradient in
the layer when the thickness of the layer is decreased
[21-28]. As a result, the dissipation in a unit volume of the
fluid, which is proportional to the square of the velocity
gradient, must also increase substantially. On the other
hand, decreasing layer thickness results in decreasing
volume in which the dissipation takes place. Thus the
overall dissipation must have a tendency to decrease with
decreasing layer thickness. The competition of these two
opposing mechanisms results in an optimal layer
thickness at which the absorptivity is maximal [21-28].

   The theoretical analysis of AA has been carried out
for both longitudinal [24,27] and shear [21-23,25,26]
bulk acoustic waves. Coefficients of absorption,
reflection, transmission and transformation have been
analysed as functions of angle of incidence, fluid layer
thickness, frequency and parameters of the media in
contact. Unusually strong absorption at the optimal
layer thickness has been predicted for all types of
acoustic waves [21-28]. The frictional contact
approximation, which in many cases allows accurate analytical
solution of the problem, has been introduced and
justified [21-26]. This approximation is equivalent to the
spring boundary conditions [1-5], but with the zero
inertial load at the contact and imaginary (for Newtonian
fluid) or complex (for non-Newtonian fluid) spring
constants for tangential displacements. Optimisation of
structural parameters for maximal absorptivity in the
layer has been carried out. The effects of non-Newtonian
properties of the fluid on AA has been investigated for
longitudinal waves [27] and bulk shear waves polarized
normally to the plane of incidence [26]. More recently,
experimental investigation of AA, and verification of the
theoretical predictions for bulk shear waves at normal
incidence have been carried out [28].

   However, although the analysis of AA of bulk
acoustic waves has been quite thorough, there has been
little attention paid to AA of shear sagittal acoustic
waves (i.e., shear waves polarized in the plane of
incidence). To our knowledge, only one paper has made an
attempt of the theoretical analysis of AA of shear
sagittal waves [23]. In addition, this paper was only
confined to two special cases when a Newtonian fluid layer
is placed either between two identical elastic media, or
between two media, one of which can be regarded as
infinitely rigid [23]. At the same time, in practice, we
often come across situations where a thin fluid layer is
enclosed between two similar but different elastic
materials (e.g., in the case of a frictional contact between
different objects). Moreover, it can be expected that AA
of shear sagittal waves in the vicinity of critical angles of
incidence will result in interesting effects that are similar
to those for longitudinal waves and shear waves
polarized perpendicular to the plane of incidence [21,24,
26,27]. In addition, AA of shear sagittal acoustic waves
is anticipated to be affected by non-Newtonian
properties of fluids at high frequencies. Neither of these
questions has been investigated theoretically to date.

   Therefore, the aim of this paper is to investigate
theoretically AA of bulk shear sagittal acoustic waves by a
layer of Newtonian or non-Newtonian fluid between two
different elastic media. The optimal layer thickness and
the coefficients of reflection, transmission, transformation
and absorption will be analysed numerically as
functions of incidence angle, frequency and parameters
of the media in contact. Significant and unusual structure
of the dependencies of absorptivity on the angle of incidence
near three critical angles is predicted. Comparison
with the case of identical elastic media [23] is carried out.

\section{Basic equations and conditions}

\begin{figure}[htb]
\centerline{\includegraphics[width=0.5\columnwidth]{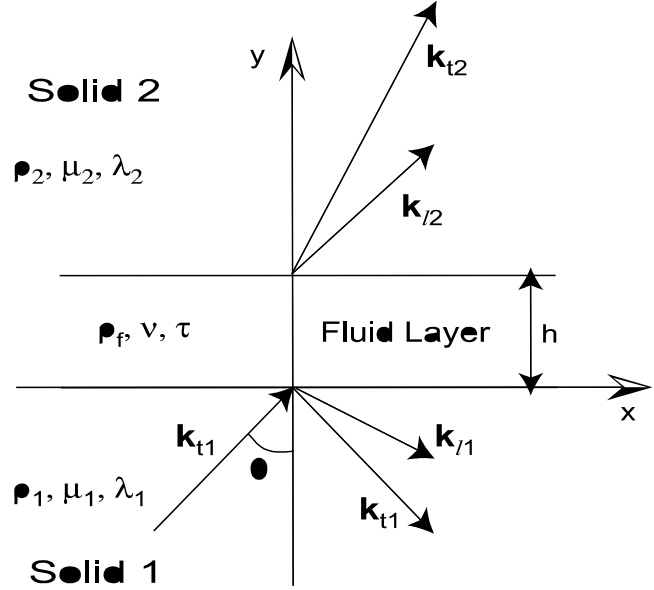}}
\caption{The structure with the anomalous absorption of bulk sagittal
shear waves in a fluid layer between two elastic media.}
\end{figure}

The analysed structure with a fluid layer enclosed
between two elastic halfspaces, is presented in Fig. 1. A
bulk shear sagittal wave propagating in the first elastic
medium is incident onto the fluid layer at an angle $\theta$. As
a result, in the first halfspace there are two reflected
waves, one of which is a shear sagittal wave and the
other is a longitudinal wave. Similarly, in the second
elastic medium there are two transmitted waves, one of
which is a shear sagittal wave and the other is a longi-
tudinal wave (Fig. 1). The elastic media 1 and 2 are
characterized by the Lame coefficients $\lambda_{1,2}$, $\mu_{1,2}$ and
densities $\rho_{1,2}$, respectively.
The vectors $\mathbf{k}_{\mathrm{l}1,2}$ are the wave
vectors of the longitudinal waves
($k_{\mathrm{l}1,2} = \omega [ \rho_{1,2}/(\lambda_{1,2}
+ 2\mu_{1,2})]^{1/2}$), and $\mathbf{k}_{\mathrm{t}1,2}$
are the wave vectors of the shear
waves ($k_{\mathrm{t}1,2} = \omega[\rho_{1,2}/\mu{1,2}]^{1/2}$)
in elastic media 1 and 2,
respectively, $\omega$ is the angular frequency. The fluid layer
has the thickness $h$, kinematic viscosity $\nu$ and density
$\rho_\mathrm{f}$.
The system of coordinates is presented in Fig. 1.

   All viscous fluids are characterized by one or more
times of relaxation, $\tau_m$ ($m = 1, 2, ...$), of shear stresses
[10,11]. If the frequency $\omega \ll \tau_m^{-1}$ for all values of $m$,
then the relaxation times can be neglected and the fluid
is Newtonian. If $\omega \approx \tau_m^{-1}$ at least for one $m$, then the
fluid is characterized not only by a shear viscosity, but
also by some non-zero shear modulus [10,11,26]. Such
fluids are called non-Newtonian fluids.

  The rigorous analysis of the interaction of an incident
shear sagittal wave with a viscoelastic layer of
non-Newtonian fluid of arbitrary thickness can be carried
out using the transfer matrix approach [8,9,19].
However, as we will see below, the absorptivity in the fluid
layer is usually strong if the layer thickness, $h$, is small
compared to the penetration depth and length of the
shear acoustic wave in the fluid. If, in addition, the effect
of the mass of the fluid layer on the wave propagation is
negligible, then the transfer matrix approach results in
the following approximate boundary conditions for the
layer [5,8]:
\begin{equation}
\sigma_{1yy} = K_n (U_{1y} - U_{2y}),
\sigma_{1xy} = K_t (U_{1x} - U_{2x}),
\sigma_{1yy} = \sigma_{2yy},
\sigma_{1xy} = \sigma_{2xy},
\end{equation}
where $\sigma_{1,2ik}$ are the strain tensor components, and $U_{1,2}$
are the displacement vectors in elastic halfspaces 1 and
2, respectively, $K_t = \mathrm{i}\omega b/h$,
$K_n = C/h$ (these coefficients
also represent spring constants in the spring boundary
conditions [1-5]), $\omega$ is the frequency of the incident
wave, $C$ is the bulk modulus of the fluid ($C = v^2 \rho_\mathrm{f}$,
$v$ is the speed of longitudinal acoustic waves in the fluid),
and
\begin{equation}
b = \sum_m \frac{\mu_{\infty m}\tau_m}{\mathrm{i}\omega\tau_m-1}
\end{equation}
where the summation is taken over all different relaxation
processes in the fluid, $\mu_{\infty m}$ are the contributions to
the overall shear modulus, $\mu_\infty = \sum_m \mu_{\infty m}$,
of the fluid at
$\omega \rightarrow + \infty$ from each of the relaxation processes.

   Usually, the speed of longitudinal waves in a viscous
fluid is much larger than the speed of shear waves caused
by shear viscosity, which means that
\begin{equation}
|C/(b\omega)| \gg 1.
\end{equation}
or $|K_n/K_t| \gg 1$. Therefore, equations (1) give
$U_{1y} - U_{2y} \ll U_{1x} - U_{2x}$. and the
displacements that are normal to the
layer must be approximately the same at the layer
boundaries $y = 0$ and $y = h$ (Fig. 1). Thus the first of the
boundary conditions (1) can be reduced as $U_{1y} = U_{2y}$.
Introducing also tangential velocities of the layer
boundaries $\partial U_{1,2x}/\partial t = -\mathrm{i}\omega U_{1,2x}$, gives
\begin{equation}
\sigma_{1yy} = \sigma_{2yy}, U_{1y} = U_{2y}, \sigma_{1xy} = \sigma_{2xy},
\sigma_{1xy} = W ( \partial U_{2x}/\partial t
- \partial U_{1x}/\partial t)_{y=0}
\end{equation}
where
\begin{equation}
W = -b/h.
\end{equation}

   Boundary conditions (4) mean that the fluid layer has
been replaced by an immediate contact of two solid
halfspaces with friction, the coefficient of which is given
by Eq. (5). Indeed, normal displacements and tangential
and normal stresses are continuous across the contact
(layer), and the tangential stresses are proportional to
the relative velocity of the surfaces in contact. The
coefficient of proportionality, W , is the coefficient of
friction, which is real for Newtonian fluid layers and
complex for non-Newtonian fluids (see Eq. (5)). Note
that this procedure is similar to that used for the
approximation of a bonding layer between two elastic
media by some special boundary conditions (e.g., spring
boundary conditions), depending on the properties of
the layer [1-5,8,9,20].

   As has been mentioned above, the main applicability
condition for the frictional contact approximation
(represented by conditions (4)) to a fluid layer between
two elastic media is that the layer thickness h must be
noticeably smaller than the magnitude of the complex
penetration depth a of the shear wave into the fluid:
\begin{equation}
h|\alpha| \ll 1,
\end{equation}
where
\begin{equation}
\alpha^2 = -\mathrm{i}\omega\rho_\mathrm{f}/b +
k^2_{\mathrm{l}1}\sin^2\theta.
\end{equation}

   However, conditions (3) and (6) are necessary but not
sufficient for the validity of boundary conditions (1) and
(4). It is also necessary to assume that the mass of the
layer can be neglected [8,9]. This can be done if, in
addition to conditions (3) and (6), the shear impedance of
the fluid is much smaller than the shear impedances of
the surrounding elastic media:
\begin{equation}
|b\alpha| \ll (\rho_1\mu_1)^{1/2}, (\rho_2\mu_2)^{1/2} 
\end{equation}
(compare with the similar conditions for the frictional
contact approximation for shear acoustic waves
polarized perpendicular to the plane of incidence in a thin
layer of Newtonian and non-Newtonian fluids [21,26]).

   Note that at lower frequencies ($\ll 100$\,MHz)
conditions (3) and (8) can be breached only if the viscosity,
density, and/or relaxation times of the fluid are unusually
large, or the shear modulus of at least one of the
surrounding elastic media is unusually small (rubber-like
material). For commonly used fluids and solids at
not very high frequencies, conditions (3) and (8) can
easily be satisfied and the main condition for the
frictional contact approximation is given by inequality (6).

   Substituting the solutions to the wave equations in
elastic media in the form of one incident, two reflected
and two transmitted waves (see Fig. 1) into boundary
conditions (4), we obtain a set of four linear
algebraic equations with four unknown wave amplitudes.
The energy coefficients of reflection, transmission and
transformation were determined as the ratios of the
$y$-components of the Poynting vectors in the relevant
reflected or transmitted waves to the $y$-component of the
Poynting vector in the incident wave. In this case the
wave absorptivity in the layer is given by:
\begin{equation}
M = 1 - T- R - K_1 - K_2,
\end{equation}
where $T$ is the transmissivity, $R$ is the reflectivity, $K_1$ is
the coefficient of transformation of the incident transverse
wave into a reflected longitudinal wave in the first
elastic medium, and $K_2$ is the coefficient of
transformation in the second elastic medium.

   One of the significant advantages of the presented
frictional contact approximation is that it allows simple
analysis of AA in non-uniform thin fluid layers. If fluid
parameters such as viscosity, density, relaxation time(s)
vary across the layer, for example due to interaction of
the fluid with the solid surfaces, then the parameter $b$
determined by Eq. (2) is a function of the $y$-coordinate
(Fig. 1). As shown in paper [26], the frictional coefficient
for such a non-uniform layer of non-Newtonian fluid of
thickness $h$ is given by the equation:
\begin{equation}
W = - \left( \int_0^h b^{-1} \mathrm{d}y \right)^{-1}.
\end{equation}

In this case condition (6) has to be written as [26]:
\begin{equation}
\int_0^h |\alpha| \mathrm{d}x \ll 1
\end{equation}
and in conditions (3) and (8) $b$ must be replaced by
$\mathrm{max}(|b|)$.

   In the frictional contact approximation, the coefficient
of friction is the only parameter that represents properties
of the fluid layer. Therefore, if we use Eq. (10) for
the frictional coefficient, then all other equations determining
AA in a non-uniform layer will remain the same
as for uniform layers.

\section{Numerical analysis}

   An attempt to carry out analytical analysis of AA of
shear sagittal waves was made in paper [23] for the
special case of a uniform Newtonian fluid layer between
two identical elastic media. However, analytical analysis
becomes much more involved if more complicated
structures are considered and, therefore, numerical
analysis of AA has been used in this paper. In addition,
since in the frictional contact approximation non-uniform
fluid layers can be analysed in the same way as
uniform layers (with the use of Eq. (10)), only examples
with uniform fluid layers will be considered in this section.

  As an example, consider a structure with a thin uni-
form layer of poly-1-butene-16 (with 16 repeat units and
average molecular weight 640 [10]), enclosed between
two solid halfspaces. The parameters of the fluid layer
are $\rho_\mathrm{f} \approx 0.88$\,g/cm$^3$,
$\nu \approx 57$\,cm$^2$/s, and $\tau \approx 6.33 \times 10^{-9}$s
[10] (only one relaxation time is taken into account,
$\mu_\infty = \rho_\mathrm{f}\nu/\tau$).

  If the first elastic medium is made of silica
($\rho_1 = 2.2$\,g/cm$^3$, $\mu_1 = 3.11 \times 10^{11}$dyne/cm$^2$),
and the second elastic medium is made of tin
($\rho_2 = 7.3$\,g/cm$^3$, $\mu_2 = 2.11\times 10^{11}$dyne/cm$^2$),
then the structure will be called silica--poly-1-butene-16--tin.
Similarly, if the first medium is tin
and the second is silica, then we will have a
tin--poly-1-butene-16--silica structure, etc.

\begin{figure}[htb]
\centerline{\includegraphics[width=0.7\columnwidth]{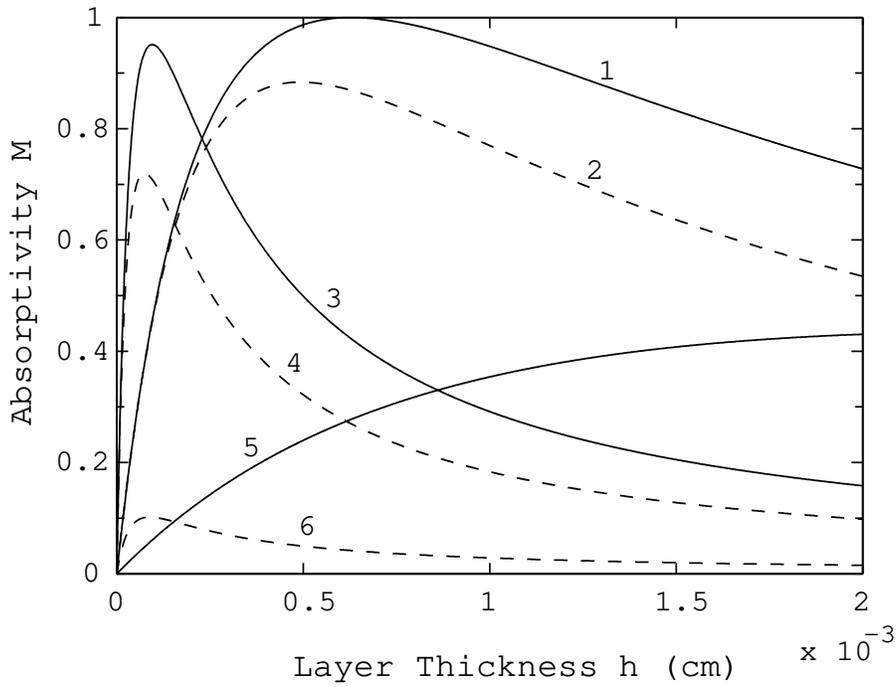}}
\caption{The dependencies of the absorptivity on layer thickness for two
different structures: tin-fluid-silica (curves 1-4), and silica-fluid-tin
(curves 5 and 6). The fluid layer: poly-1-butene-16 with
$\rho_\mathrm{f} = 0.88$\,g/cm$^3$, $\nu = 57$\,cm$^2$/s,
$\tau = 6.33 \times 10^{-9}$s [10] and $\omega/2\pi = 20$\,MHz (curves 2
and 4); the Newtonian fluid with the same $\rho_\mathrm{f}$ and
$\nu$ but $\omega\tau \ll 1$ (curves
1, 3, 5, and 6). Angles of incidence
$\theta = \theta_\mathrm{cl}(\mathrm{tin}) \approx 38.504^\circ$
(curves 1 and 2), $\theta = 70^\circ$ (curves 3, 4, and 6),
$\theta = \theta_\mathrm{cl}(\mathrm{silica}) \approx 39.046^\circ$
(curve 5).}
\end{figure}

   Fig. 2 presents typical dependencies of the absorp-
tivity $M$ on layer thickness $h$ for two different structures:
tin--fluid--silica (curves 1-4) and silica--fluid--tin (curves 5
and 6). The angles of incidence are
$\theta = \theta_\mathrm{cl}(\mathrm{tin}) \approx 38.504^\circ$
(curves 1 and 2), $\theta = 70^\circ$ (curves 3, 4, and 6),
and $\theta = \theta_\mathrm{cl}(\mathrm{silica}) \approx 39.046^\circ$
(curve 5), where $\theta_\mathrm{cl}$ is the
critical angle at which the reflected longitudinal wave in
medium 1 propagates parallel to the layer. Curves 2
and 4 are presented for the non-Newtonian fluid layer
made of poly-1-butene-16 with $\tau = 6.33 \times 10^{-9}$s at the
acoustic frequency $\omega/2\pi = 20$\,MHz ($\omega\tau \approx 0.8$), while
curves 1, 3, 5, and 6 are presented for the Newtonian
fluid layer with the same $\rho_\mathrm{f}$ and $\nu$ as poly-1-butene-16,
but at $\omega\tau\ll 1$. Note that in the frictional contact
approximation all curves for Newtonian fluid layers (e.g.,
curves 1, 3, 5, and 6 in Fig. 2) are frequency
independent---see also [21,23,24,26,27].

   All the curves in Fig. 2 display the behaviour typical
of AA (see also [21-27]). The absorptivity strongly
increases with decreasing layer thickness, reaches a
relatively broad maximum at an optimal thickness, $h_m$, and
then quickly goes to zero as the layer thickness tends to
zero (Fig. 2). Note that for curve 5 the absorptivity
maximum is reached at the optimal thickness $h_m \approx 25\mu$m
and therefore cannot be seen in Fig. 2. The maximums
of the absorptivity dependencies are broad, which
reflects the non-resonant character of AA (see also [21-26]).
The absorptivity curves for a non-Newtonian fluid
are lower than for the corresponding Newtonian fluid
(compare curves 1, 3 and 2, 4 in Fig. 2). This is the result
of the general tendency: the larger the product $\omega\tau$, the
smaller the absorptivity maximum due to AA. Note
however that the difference between the curves for the
Newtonian and non-Newtonian fluids is maximal at
layer thicknesses near (or a few times larger than) the
optimal thickness. If $h \ll h_m$ or $h \gg h_m$, the sensitivity
of the absorptivity to non-zero relaxation time becomes
noticeably weaker. That is why AA may especially be
useful for the analysis of non-Newtonian properties of
fluids (see also [28]).

  Comparison of curves 1-4 and 5, 6 in Fig. 2 suggests
that it is preferable to use structures with the first elastic
medium having smaller speeds of acoustic waves than in
the second medium. In this case stronger AA can be
achieved (Fig. 2).

\begin{figure}[htb]
\centerline{\includegraphics[width=0.7\columnwidth]{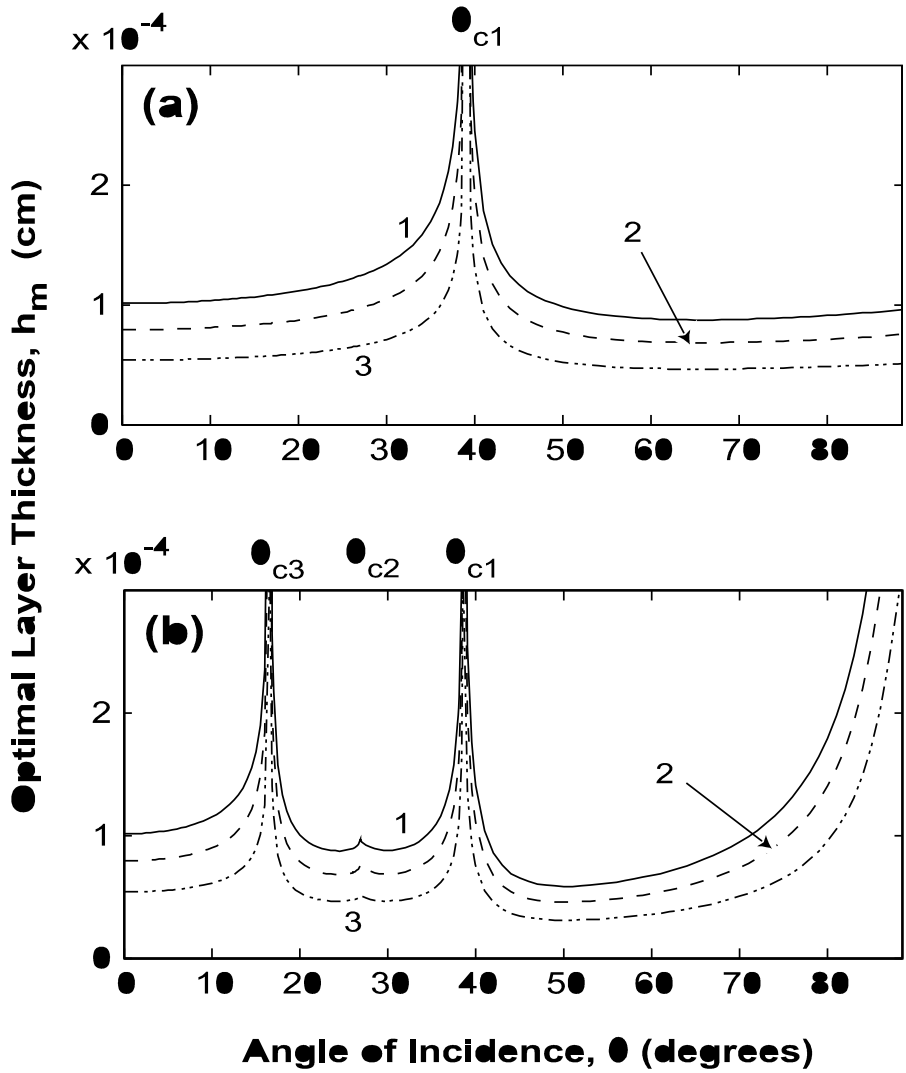}}
\caption{The angular dependencies of the optimal layer thickness for the
silica-fluid-tin structure (a), and tin-fluid-silica structure (b). The fluid
layer: (1) poly-1-butene-16 with xs ( 1 (Newtonian fluid); (2)
poly-1-butene-16 with
$\tau = 6.33 \times 10^{-9}$s and $\omega/2\pi = 20$\,MHz;
(3) same as (2) but with $\omega/2\pi = 40$\,MHz.}
\end{figure}

   The angular dependencies of the optimal layer thickness,
$h_m(\theta)$, for the silica--poly-1-butene-16--tin structure
are presented in Fig. 3a. Curve 1 in Fig. 3a is for
the poly-1-butene-16 layer with the assumption that
$\omega\tau\ll 1$, while curves 2 and 3 are for the cases with
$\tau = 6.33 \times 10^{-9}$s and the frequencies $\omega/2\pi = 20$
and 40\,MHz, respectively. All three curves are characterized by
strong maximums at the critical angle
$\theta_\mathrm{cl} \approx 39.046^\circ$.
The values of these maximums are 25, 20, and 14\,$\mu$m for
curves 1, 2, and 3, respectively. It can be seen that the
effect of non-Newtonian properties of the fluid is the
reduction in the optimal layer thickness (Fig. 3a).

   Fig. 3b displays the dependencies $h_m(\theta)$ for the
tin--poly-1-butene-16--silica structure for the frequencies
$\omega/2\pi = 20$\,MHz (curve 2) and
$\omega/2\pi = 40$\,MHz (curve 3).
Curve 1 represents the same structure, but with
$\omega\tau\ll 1$ (i.e., for the Newtonian fluid layer). The three
maximums of each of these dependencies correspond to
three critical angles of incidence in the considered
structure. As has been mentioned, the first critical angle,
$\theta_\mathrm{c1} \approx 38.504^\circ$ corresponds to the
reflected longitudinal
wave in medium 1 (tin) propagating parallel to the layer
($\mathbf{k}_x = \mathbf{k}_{\mathrm{t}1x} = \mathbf{k}_{\mathrm{l}1}$).
The second critical angle
$\theta_\mathrm{c2} \approx 26.9^\circ$
corresponds to the transmitted shear wave propagating
parallel to the layer ($\mathbf{k}_x = \mathbf{k}_{\mathrm{t}2}$).
Finally, the third critical
angle $\theta_\mathrm{c2} \approx 16.55^\circ$
corresponds to the longitudinal wave
in medium 2 propagating parallel to the layer (this
happens when $\mathbf{k}_x = \mathbf{k}_{\mathrm{l}2}$).
If $\theta = \theta_{\mathrm{c}1}$, then the optimal
layer thickness for the Newtonian fluid (curve 1) is
6.9\,$\mu$m, while for curves 2 and 3 it is equal to 5.4 and
3.9\,$\mu$m, respectively. Similarly, if
$\theta = \theta_{\mathrm{c}3}$, then the optimal
layer thickness for the Newtonian fluid reaches 25\,$\mu$m,
while for curves 2 and 3 it is 20 and 14\,$\mu$m. At the second
critical angle the optimal layer thicknesses are characterized
by only a very minor peak (Fig. 3b).

  The curves in Fig. 3b also demonstrate a reduction in
the optimal layer thicknesses with increase of fluid
relaxation time or frequency. This is similar to the
tendency that has been observed in Fig. 3a. Note again that
curves 1 for Newtonian fluid layers in Fig. 3a and b are
independent of frequency.

   The dependencies $h_m(\theta)$ in Fig. 3a are typical of
structures where the speed of transverse waves in medium
1 is larger than the speed of longitudinal waves in
medium 2 ($c_{\mathrm{t}1} > c_{\mathrm{l}2}$),
whereas the dependencies in Fig.
3b are typical of structures where $c_{\mathrm{t}1} < c_{\mathrm{t}2}$.

\begin{figure}[htb]
\centerline{\includegraphics[width=0.7\columnwidth]{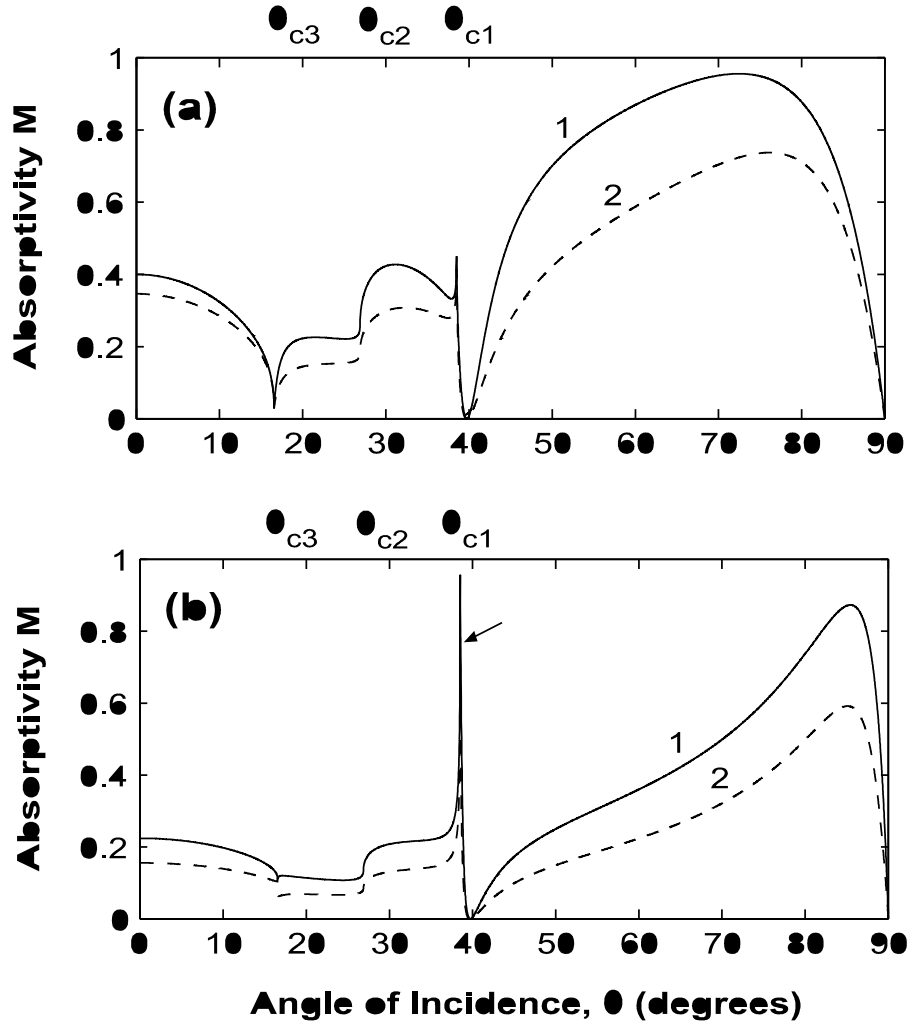}}
\caption{The angular dependencies of the absorptivity $M$ for the
tin--fluid--silica structure for the two layer thicknesses:
(a) $h = 1\mu$, (b) $h = 5\mu$m. The fluid layer: poly-1-butene-16
with $\omega\tau\ll 1$ (Newtonian fluid---curves 1);
and poly-1-butene-16 with $\tau = 6.33 \times 10^{-9}$s and
$\omega/2\pi = 20$\,MHz (curves 2). The arrow indicates the maximal value of
the absorptivity for curve 2.}
\end{figure}

   The dependencies of the absorptivity on angle of
incidence for the tin--fluid--silica structure are presented in
Fig. 4a and b for two different layer thicknesses: (a)
$h = 1\mu$m, and (b) $h = 5\mu$m. Curves 1 are again for the
Newtonian layer (i.e., poly-1-butene-16 with $\omega\tau\ll 1$),
while curves 2 are for the non-Newtonian fluid layer
with $\tau = 6.33 \times 10^{-9}$s and $\omega/2\pi = 20$.

   The most interesting feature of Fig. 4a and b is the
sharp and strong maximum of absorptivity at the first
critical angle (Fig. 4b) for both Newtonian and
non-Newtonian fluids. It is important that this maximum is
achieved at the same angle at which the optimal layer
thickness also has a strong maximum (Fig. 3b). This
behaviour is unique to AA of sagittal shear waves, since
the usual tendency is that increasing optimal layer
thickness (e.g., near the critical angle for shear waves
polarized normally to the plane of incidence) results in a
simultaneous decrease of the absorptivity maximum
[21,22,26]. The coincidence of the maximums of the
absorptivity and optimal layer thickness (Figs. 3b
and 4b) may also be important for observation and
application of AA in fluid layers with small viscosity,
where optimal thickness may be unreasonably small
[21,22,26].

\begin{figure}[htb]
\centerline{\includegraphics[width=0.7\columnwidth]{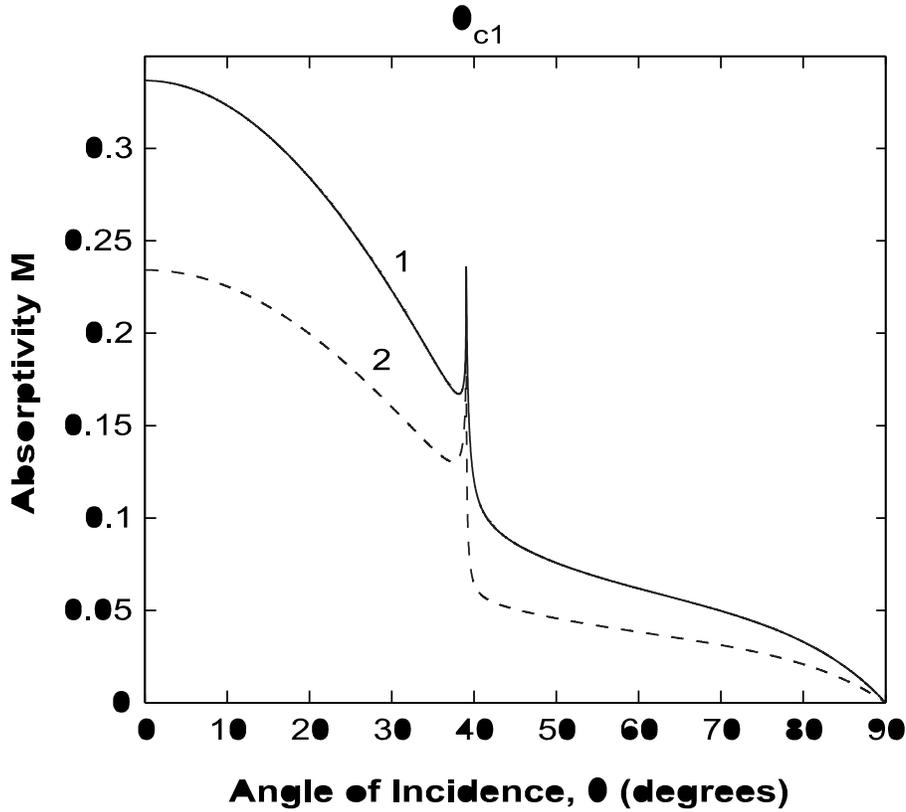}}
\caption{The angular dependencies of the absorptivity $M$ for the
tin--fluid--silica structure for the layer thickness $h = 5\mu$m.
The fluid layer: (1) poly-1-butene-16 with $\omega\tau\ll 1$
(Newtonian fluid); (2) poly-1-butene-16 with
$\tau = 6.33 \times 10^{-9}$ and $\omega/2\pi = 20$MHz.}
\end{figure}

   Figs. 3b and 4a,b are typical for the case when the
speed of longitudinal acoustic waves in medium 1,
$c_{\mathrm{l}1}$, is
less than the speed of transverse waves in medium 2,
$c_{\mathrm{t}2}$,
i.e., when three critical angles exist and
$\theta_{\mathrm{c}2,3} < \theta_{\mathrm{c}1}$. If
only one critical angle exists, i.e., if the speed of
longitudinal waves in medium 2, $c_{\mathrm{l}2}$,
is smaller than the speed
of the incident transverse wave, $c_{\mathrm{t}1}$,
(e.g., for the silica--fluid--tin structure), then the
typical angular dependencies of the absorptivity are
presented in Fig. 5 for the
layer thickness $h = 5\mu$m for Newtonian (curve 1) and
non-Newtonian (curve 2) fluids. It can be seen that in
this case the overall absorptivity is significantly smaller.
However, a sharp (but smaller) maximum at the first
(and the only) critical angle still exists for both Newtonian
and non-Newtonian fluid layers (Fig. 5).

One may think that the sharp maximums of the angular
dependencies of the absorptivity at the critical
angles (Figs. 4 and 5) may have the same physical
explanation as the maximums of conversion of a sagittal
shear wave into a longitudinal wave at an interface
between two media due to the generation of
pseudo-surface-waves at the interface [20]. However, in our
opinion, this interpretation is questionable. Indeed, the
maximums of the longitudinal wave amplitudes in [20]
occur at angles that are slightly (but noticeably) larger
than the critical angle, whereas all the maximums in
Figs. 3--5 and below occur precisely at the critical angles.
Furthermore, it can be seen that the absorptivity due to
AA at the angle corresponding to the maximal wave
transformation [20] is substantially lower than its
maximum value at the critical angle (Figs. 4 and 5). For
example, for the sapphire--copper structure considered
in [20] the maximum of the absorptivity due to AA (if
there is a fluid layer between the elastic media) and the
maximum of the wave transformation [20] hardly overlap.
This suggests different physical nature of these
maximums and effects.

   At the same time, the sharp features of the angular
dependencies of the reflected and refracted waves at the
critical angles at a sliding contact between two elastic
media [7] obviously have a relationship to the behaviour
of the similar dependencies in the case of AA in a fluid
layer at the same angles. However, it is necessary to
stress out that AA does not exist on a sliding contact,
because such a contact is non-dissipative (without
friction) [7].

\begin{figure}[htb]
\centerline{\includegraphics[width=0.7\columnwidth]{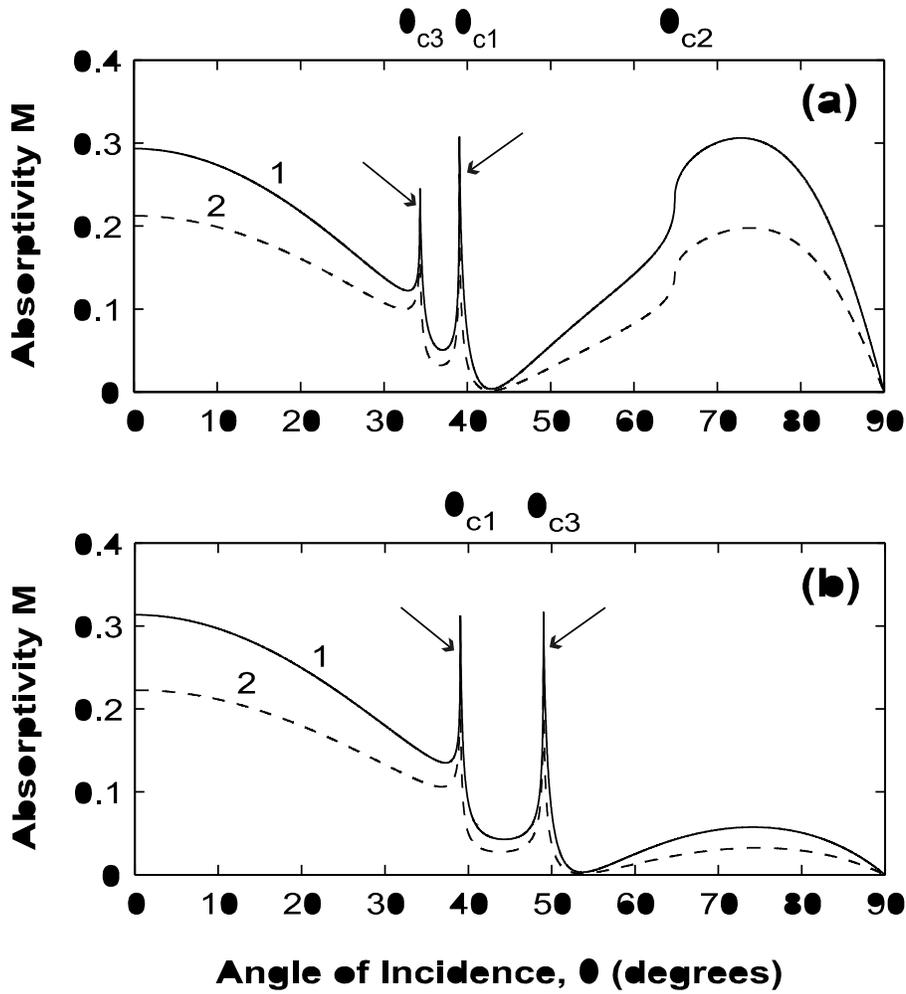}}
\caption{The angular dependencies of the absorptivity $M$ in the structure
silica--fluid--medium 2, where the parameters of the hypothetical
medium 2 are: (a) $\rho_2 = 1.45$\,g/cm$^3$,
$\mu_2 = 2.5\times 10^{11}$dyne/cm$^2$,
$\lambda_2 = 1.45 \times 10^{11}$dyne/cm$^2$; (b) $\rho_2 = 2.6$\,g/cm$^3$
and $\mu_2$ and $\lambda_2$ are the
same as in (a). The layer thickness is $h = 5\mu$m. The fluid layer:
poly-1-butene-16 with $\omega\tau\ll 1$ (Newtonian fluid---curves 1);
and poly-1-butene-16 with
$\tau = 6.33 \times 10^{-9}$ and $\omega/2\pi = 20$MHz (curves 2). The
arrows indicate the maximal values of the absorptivity for curves 2.}
\end{figure}

   Fig. 6a and b present typical angular dependencies of
the absorptivity for structures where
$c_{\mathrm{t}1} < c_{\mathrm{t}2}$ and
$c_{\mathrm{t}2} < c_{\mathrm{l}1} < c_{\mathrm{l}2}$
(i.e., $\theta_{\mathrm{c}3} < \theta_{\mathrm{c}1}
< \theta_{\mathrm{c}2}$; $\theta_{\mathrm{c}1} \approx 39.046^\circ$,
$\theta_{\mathrm{c}2} \approx 64.9^\circ$,
$\theta_{\mathrm{c}3} \approx 34.314^\circ$---Fig. 6a),
and $c_{\mathrm{t}1} < c_{\mathrm{l}2} < c_{\mathrm{l}1}$ but
$c_{\mathrm{t}1} > c_{\mathrm{t}2}$, i.e.,
$\theta_{\mathrm{c}3} > \theta_{\mathrm{c}1}$ and
$\theta_{\mathrm{c}2}$ does not exist
($\theta_{\mathrm{c}1} \approx 39.046^\circ$,
$\theta_{\mathrm{c}3} \approx 49.015^=\circ$---Fig. 6b).
It can be seen that in
these cases the absorptivity is characterized by two
sharp and strong maximums---at $\theta_{\mathrm{c}1}$ (as in Figs. 4 and
5), and at $\theta_{\mathrm{c}3}$. Note however, that these maximums are
not as strong as the maximum in Fig. 4b. In the
tin--fluid--silica structure the absorptivity also experiences a
sharp maximum at the angle $\theta = \theta_{\mathrm{c}3}$,
but at significantly
larger layer thicknesses (e.g., at $h = 20\mu$m) than those
used for plotting Fig. 4a and b. This is the reason for not
observing this maximum at
$\theta = \theta_{\mathrm{c}3} \approx 16.55^\circ$ in Fig. 4a
and b. Thus, similar to the absorptivity maximum at
$\theta = \theta_{\mathrm{c}1}$, the maximum at
$\theta = \theta_{\mathrm{c}3}$ also coincides with a
strong maximum in the optimal layer thickness (see Fig. 3b).

Note that the absorptivity maximum at the first critical
angle has also been predicted for a Newtonian fluid
layer between two identical elastic media [23]. However,
in that case, the absorptivity maximum was restricted to
50\% [23], whereas in the case of different media it can go
as high as 100\%---Figs. 2 and 4b. In addition, in the case
of different elastic media we can obtain two sharp
maximums at the first and the third critical angles---Fig.
6a and b. Actually, for identical elastic media these two
maximums coincide with each other, giving only one
maximum that can reach 50\% [23]. Therefore, an interesting
question is what will happen to this maximum if
the parameters of the two elastic media are only slightly
different? How sensitive is AA to small variations of the
elastic media? These questions are especially interesting
due to the sharpness of the absorptivity maximum at
$\theta = \theta_{\mathrm{c}1}$,
and due to small natural fluctuations of parameters of elastic media.

\begin{figure}[htb]
\centerline{\includegraphics[width=0.7\columnwidth]{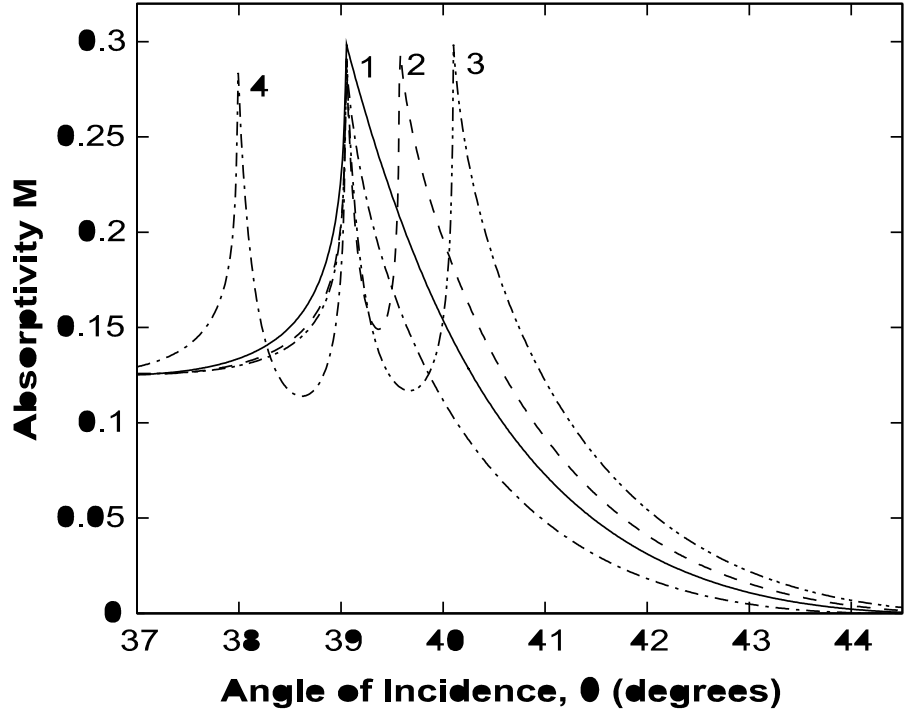}}
\caption{The angular dependencies of the absorptivity $M$ in the structure
silica--fluid--silica, where the density of the second silica halfspace
$\rho_2$ is
slightly different from the first halfspace:
(1) $\rho_2 = \rho_1 = 2.2$\,g/cm$^3$,
(2) $\rho_2 = 2.25$\,g/cm$^3$ and $\rho_1 = 2.2$\,g/cm$^3$,
(3) $\rho_2 = 2.3$\,g/cm$^3$ and $\rho_1 = 2.2$\,g/cm$^3$,
(4) $\rho_2 = 2.1$\,g/cm$^3$ and $\rho_1 = 2.2$\,g/cm$^3$.
The layer thickness is $h = 5\mu$m.}
\end{figure}

   Therefore, Fig. 7 presents the angular dependencies of
the absorptivity in a layer of poly-1-butene-16 (at
$\omega\tau\ll 1$ of thickness $h = 5\mu$m between two silica
halfspaces with small variations of the density of the second
silica halfspace: (1) $\rho_2 = \rho_1 = 2.2$\,g/cm$^3$,
(2) $\rho_2 = 2.25$\,g/cm$^3$ and $\rho_1 = 2.2$\,g/cm$^3$,
(3) $\rho_2 = 2.3$\,g/cm$^3$ and $\rho_1 = 2.2$\,g/cm$^3$,
(4) $\rho_2 = 2.1$\,g/cm$^3$ and $\rho_1 = 2.2$\,g/cm$^3$.
The elastic
modulae of both the silica halfspaces are assumed to be
the same: $\mu_1 = \mu_2 = 3.11 \times 10^{11}$dyne/cm$^2$,
$\lambda_1 = \lambda_2 = 1.62 \times 10^{11}$dyne/cm$^2$.
It can be seen that in the case of
different densities $\rho_1 \ne \rho_2$ (curves 2--4 in Fig. 7) the
single maximum (curve 1) splits into two similar sharp
maximums. The angular distance between these maximums
increases with increasing difference between the
densities of the elastic halfspaces, and the dip between
the maximums becomes more pronounced (Fig. 7).

\begin{figure}[htb]
\centerline{\includegraphics[width=0.7\columnwidth]{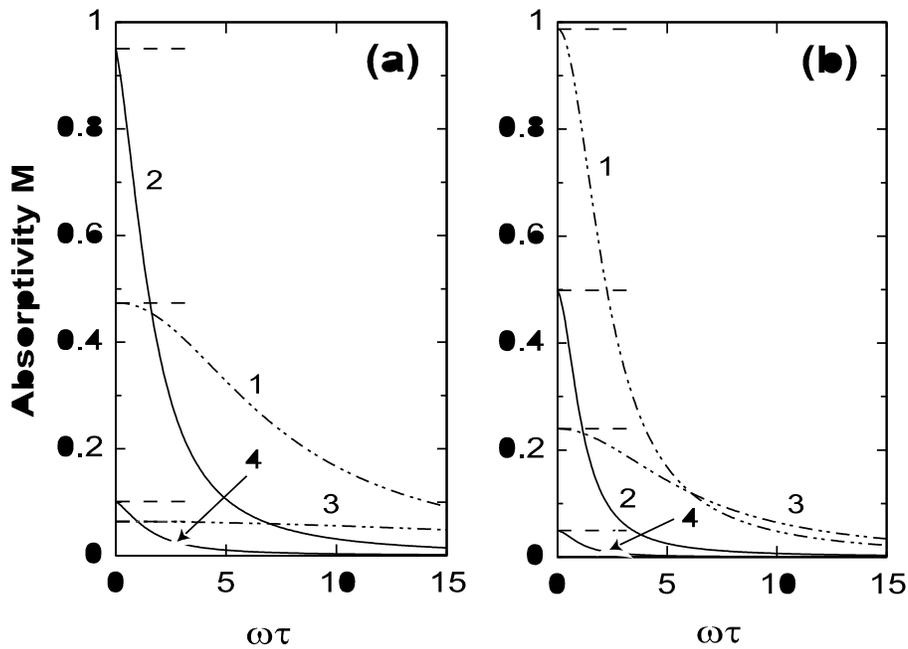}}
\caption{The dependencies of the absorptivity $M$ on the product
$\omega\tau$ for
two different thicknesses of the layer:
(a) $h = 1\mu$m, (b) $h = 5\mu$m.
Curves 1: tin--poly-1-butene-16--silica,
$\theta = \theta_{\mathrm{c}1}(\mathrm{tin}) \approx 38.504^\circ$.
Curves 2: tin--poly-1-butene-16--silica, $\theta = 70^\circ$.
Curves 3: silica--poly-1-butene-16--tin,
$\theta = \theta_{\mathrm{c}1}(\mathrm{silica}) \approx 39.046^\circ$.
Curves 4: silica--poly-1-butene-16--tin, $\theta = 70^\circ$.
The dashed straight lines indicate the absorptivities for the
corresponding Newtonian fluid layers (with the same as for
poly-1-butene-16 viscosity and density, but with $\omega\tau\ll 1$}
\end{figure}

   Finally, Fig. 8a and b present the frequency
dependencies of the absorptivity maximum for the
tin--poly-1-butene-16--silica (curves 1 and 2) and
silica--poly-1-butene-16--tin (curves 3 and 4) structures for two
different layer thicknesses $h = 1\mu$m (Fig. 8a) and $h = 5\mu$
(Fig. 8b) in the approximation of the frictional
contact approximation. The short dashed lines parallel
to the horizontal axis correspond to the Newtonian fluid
layers with the same thicknesses, viscosity and density.

   The main aspect that can be seen from Fig. 8a and b is
that the non-Newtonian properties of the fluid produce
significantly stronger effect on AA of sagittal shear
waves if the absorptivity is large. As a result, the
sensitivity of AA-based viscosity sensors to non-Newtonian
properties of fluids must be substantially higher than of
those using absorption of acoustic waves at an isolated
solid--fluid interface [10--16]. This important feature of
AA has also been investigated experimentally in paper
[28] for the normally incident shear acoustic waves. This
is the main reason why AA is of a special importance for
the experimental investigation of non-Newtonian properties
of fluids [28].

   The main physical reason for the strong sensitivity of
AA to non-zero relaxation time(s) in the fluid is related
to the fact that due to strong absorption in the layer, the
range of variations of the absorptivity is much larger
(see curves 1 and 2 in Fig. 8a and b) than for an isolated
solid--fluid interface [10,11]. Therefore, as was mentioned
above during the discussion of Fig. 2, it is beneficial to
use the structures with the second elastic
medium being more rigid than the first, since in this case
the absorptivity can be especially large---see Figs. 2 and
8a,b.

\section{Conclusions}

   This paper has analysed the main features of the
anomalous absorption of bulk shear sagittal waves in
layers of Newtonian and non-Newtonian fluids between
two different elastic media in the frictional contact
approximation. The absorptivity in an ultra-thin fluid
layer of thickness, that is much smaller than the wavelength
and penetration depths of longitudinal and shear
acoustic waves in the fluid, has been demonstrated to be
much larger than the absorptivity of bulk acoustic waves
at an isolated solid­fluid interface. An interesting
behaviour that is characterized by pronounced maximums
in the angular dependencies of the absorptivity has been
demonstrated and investigated numerically for various
structural parameters including layer thickness. In
particular, it has been shown that strong and sharp
absorptivity maximums occur at two critical angles of
incidence at which the reflected and transmitted longitudinal
waves in the surrounding elastic media propagate
parallel to the layer. An important and unique
feature of AA of sagittal shear waves is that these two
strong maximums of the absorptivity coincide with
simultaneous strong maximums of the optimal layer
thickness. This fact may be important for the experimental
observation and applications of AA for the
analysis of fluid layers with small viscosity, where the
optimal layer thicknesses may be unreasonably small.
The effect of small variations in parameters of the elastic
media on the absorptivity maximums has been analysed.

   It has been demonstrated theoretically that one of the
most important features of AA is its much higher
sensitivity to non-Newtonian properties of fluids (i.e.,
non-zero times(s) of relaxation of shear stresses in viscous
fluids) than that of the absorption of acoustic waves at
an isolated solid--fluid interface. The stronger the
anomalous absorption, the stronger its sensitivity to
non-Newtonian properties of the fluid layer. This makes
AA especially important for the experimental investigation
of non-Newtonian properties of fluids and particularly
solid--fluid interfaces.

   The developed frictional contact approximation
has been shown to be very useful for simple analysis of
AA in non-uniform fluid layers with fluid parameters
(i.e., density, viscosity, and/or relaxation times) varying
across the layer. This is important for the analysis of
interactions (e.g., adsorption--desorption processes) at
solid--fluid interfaces, resulting in stratification of the
fluid in the vicinity of a solid surface.

   As a result, feasible applications of AA of acoustic
waves can be anticipated in such areas as investigation
and diagnostics of ultra-small amounts of fluids
(less than $\approx 10^{-5}$cm$^3$), suspensions, colloidal
solutions, biological and human fluids, solid­fluid interfaces,
frictional contacts and lubricants in tribology,
non-destructive evaluation, etc.

\section*{Acknowledgements}

  The authors gratefully acknowledge financial support
for this work from the Queensland University of Technology.

\section*{References}

\begin{enumerate}
\item J.-M. Baik, R.B. Thompson, Ultrasonic scattering from imperfect
    interfaces: a quasi-static model, J. Nondestruct. Eval. 4 (1984)
    177--196.
\item F.J. Margetan, R.B. Thompson, T.A. Gray, Interfacial spring
    model for ultrasonic interactions with imperfect interfaces: theory
    of oblique incidence and application to diffusion-bonded butt
    joints, J. Nondestruct. Eval. 7 (1988) 131--152.
\item P.B. Nagy, L. Adler, Ultrasonic NDE of solid-state bonds:
    internal and friction welds, J. Nondestruct. Eval. 7 (1988)
    199--215.
\item P.B. Nagy, Ultrasonic classification of imperfect surfaces, J.
    Nondestruct. Eval. 11 (1992) 127--139.
\item A.I. Lavrentyev, S.I. Rokhlin, Ultrasonic spectroscopy of
    imperfect contact interfaces between a layer and two solids, J. Acoust.
    Soc. Am. 103 (1998) 657--664.
\item A.L. Shuvalov, J. Lothe, The Stroh formalism and the reciprocity
    properties of reflection-transmission problems in crystal
    piezo-acoustics, Wave Motion 25 (1997) 331--345.
\item A.L. Shuvalov, A.S. Gorkunova, Cutting-off effect at reflection-%
     transmission of acoustic waves in anisotropic media with sliding-%
     contact interfaces, Wave Motion 30 (1999) 345--365.
\item S.I. Rokhlin, Y.J. Wang, Analysis of boundary conditions for
     elastic wave interaction with an interface between two solids, J.
     Acoust. Soc. Am. 89 (1991) 503--515.
\item S.I. Rokhlin, Y.J. Wang, Equivalent boundary conditions for thin
     orthotropic layer between two solids: reflection, refraction, and
     interface waves, J. Acoust. Soc. Am. 91 (1992) 1875--1887.
\item A.J. Barlow, R.A. Dickie, J. Lamb, Viscoelastic relaxation in
     poly-1butenes of low molecular weight, Proc. Roy. Soc. A 300
     (1967) 356--372.
\item R.S. Moore, H.G. McSkimin, in: Physical Acoustics, vol. 6,
     Academic Press, New York, 1970, pp. 167--243.
\item A.J. Ricco, S.J. Martin, Acoustic wave viscosity sensor, Appl.
     Phys. Lett. 50 (1987) 1474--1476.
\item T.M. Niemczyk, S.J. Martin, G.C. Frye, A.J. Ricco, Acousto-%
     electric interaction of plate modes with solutions, J. Appl. Phys. 64
     (1988) 5002--5008.
\item T. Sato, H. Okajima, Y. Kashiwase, R. Motegi, H. Nakajima,
     Shear horizontal acoustic plate mode viscosity sensor, Jpn. J.
     Appl. Phys. 32 (5B1) (1993) 2392--2395.
\item V. Shah, K. Balasubramaniam, Effect of viscosity on ultrasound
     wave reflection from a solid/liquid interface, Ultrasonics 34 (1996)
     817--824.
\item M. Thompson, G.K. Dhaliwal, C.L. Arthur, G.S. Calabreses,
     The potential of the bulk acoustic wave device as a liquid-%
     phase immunosensor, IEEE Trans. UFFC-34 (2) (1987) 127--%
     135.
\item M. Schoenberg, Elastic wave behaviour across linear slip
     interfaces, J. Acoust. Soc. Am. 68 (1980) 1516--1521.
\item O. Lenoir, J.-L. Izbicki, M. Rousseau, F. Coulouvrat, Subwavelength
     ultrasonic measurement of a very thin fluid layer thickness
     in a trilayer, Ultrasonics 35 (1997) 509.
\item L.M. Brekhovskikh, Waves in Layered Media, Academic, London, 1960.
\item A.G. Every, G.L. Koos, J.P. Wolfe, Ballistic phonon imaging in
     sapphire: bulk focusing and critical-cone channelling effects, Phys.
     Rev. B 29 (1984) 2190--2209.
\item D.K. Gramotnev, S.N. Ermoshin, L.A. Chernozatonskii, Anomalous
     absorption of shear waves in transmission across a gap filled
     with viscous fluid, Akust. Zh. 37 (1991) 660--669, Sov. Phys.
     Acoust. 37 (1991) 343--348.
\item M.L. Vyukov, D.K. Gramotnev, L.A. Chernozatonskii, Anomalous
     absorption of shear acoustic waves in a multi-layer structure
     with fluid, Akust. Zh. 37 (1991) 448--454, Sov. Phys. Acoust. 37
     (1991) 229--233.
\item L.A. Chernozatonskii, D.K. Gramotnev, M.L. Vyukov, Anomalous
     absorption of sagittal shear waves by a thin fluid layer, J.
     Physics D 25 (1992) 226--233.
\item L.A. Chernozatonskii, D.K. Gramotnev, M.L. Vyukov, Anomalous
     absorption of longitudinal acoustic waves by a thin layer of
     viscous fluid, Phys. Lett. A 164 (1992) 126--132.
\item D.K. Gramotnev, M.L. Vyukov, Anomalous absorption of shear
     plate modes by a thin layer of viscous fluid, Int. J. Mod. Phys. B 8
     (1994) 1741--1764.
\item D.K. Gramotnev, M.L. Mather, Anomalous absorption of shear
     acoustic waves by a thin layer of non-Newtonian fluid, J. Acoust.
     Soc. Am. 106 (1999) 2552--2559.
\item D.K. Gramotnev, M.L. Mather, T.A. Nieminen, Anomalous
     absorption of bulk longitudinal acoustic waves in a layered
     structure with viscous fluid, in: Proceedings of the 1998 IEEE
     Ultrasonics Symposium, 1998, pp. 1203--1206.
\item M.J. Lwin, D.K. Gramotnev, M.L. Mather, J.M. Bell, W. Scott,
     Experimental observation of anomalous absorption of bulk shear
     acoustic waves by a thin layer of viscous fluid, Appl. Phys. Lett. 76
     (2000) 2020--2022.
\end{enumerate}

\end{document}